
\newcount\notenumber

\def\note{\global\advance\notenumber by 1
\footnote{{\mathsurround=0pt$^{\the\notenumber}$}}
}

\mag=1300
\normalbaselineskip=11.66pt
\normallineskip=2pt minus 1pt
\normallineskiplimit=1pt
\normalbaselines
\vsize=17.3cm
\hsize=12.1cm
\parindent=20pt
\smallskipamount=3.6pt plus 1pt minus 9pt
\abovedisplayskip=1\normalbaselineskip plus 3pt minus 9pt
\belowdisplayskip=1\normalbaselineskip plus 3pt minus 9pt
\skip\footins=2\baselineskip
\advance\skip\footins by 3pt

\mathsurround 2pt
\newdimen\leftind \leftind=0cm
\newdimen\rightind \rightind=0.65cm

\def\pagenumbers{\footline={\hss\tenrm\folio\hss}}
\nopagenumbers




\font\seventeenbf=cmbx12 at 14pt

\def\MainHead{{\baselineskip 16.7pt\seventeenbf
\noindent\titolo\par}
\normalbaselines
\vskip 28.34pt
\noindent\autori\par
\ni\indirizzo
\footnote{\phantom{i}}{\piedipagina}
\vskip 93.34pt
\ni {\bf Abstract.} \Abstract
\np
}

\font\tenrm=cmr10
\font\tenit=cmti10
\font\tensl=cmsl10
\font\tenbf=cmbx10
\font\tentt=cmtt10

\def\tenpoint{%
\def\rm{\fam0\tenrm}%
\def\it{\fam\itfam\tenit}%
\def\sl{\fam\slfam\tensl}%
\def\tt{\fam\ttfam\tentt}%
\def\bf{\fam\bffam\tenbf}%
}

\tenpoint\rm

%
%
%
%
%
\font\ChapTitle=cmbx12
\font\SecTitle=cmbx12
\font\SubSecTitle=cmbx12
%
%
%
\def\NormalSkip{\parskip=5pt\parindent=15pt\baselineskip=12pt\PageNumbers%
\leftskip=0cm\rightskip=0cm}

\def\PageNumbers{\footline={\hss\tenrm\folio\hss}}
\def\NoPageNumbers{\footline={}}

%
%
%
\def\np{\vfill\eject}
\def\ss{\vskip 5pt}
\def\ms{\vskip 15pt}
\def\bs{\vskip 30pt}

\def\ni{\noindent}
%
%
%
\newcount\CHAPTER		          %
\newcount\SECTION		          %
\newcount\SUBSECTION		       %
\newcount\FNUMBER     
%
%
%
\CHAPTER=0		          
\SECTION=0		          
\SUBSECTION=0		       
\FNUMBER=0		          
%
%
%
\long\def\NewChapter#1{\global\advance\CHAPTER by 1%
\np\NoPageNumbers\ \vfil\parindent=0cm\leftskip=2cm\rightskip=2cm{\ChapTitle #1\hfil}%
\vskip 4cm\ \vfil\eject\SECTION=0\SUBSECTION=0\FNUMBER=0\NormalSkip}
%
%
%
\gdef\Mock{}
\long\def\NewSection#1{\bs\global\advance\SECTION by 1%
\ni{\SecTitle \ifnum\CHAPTER>0 \the\CHAPTER.\fi\the\SECTION.\ #1}\ms\SUBSECTION=0\FNUMBER=0}
\def\CurrentSection{\global\edef\Mock{\the\SECTION}} %
%
%
\long\def\NewSubSection#1{\global\advance\SUBSECTION by 1%
{\SubSecTitle \ifnum\CHAPTER>0 \the\CHAPTER.\fi\the\SECTION.\the\SUBSECTION.\ #1}\ss\FNUMBER=0}
%
%
%
%
\def\RightFormulaNumber{\global\advance\FNUMBER by 1\eqno{\fopen\the\FNUMBER\fclose}}
\def\LeftFormulaNumber{\global\advance\FNUMBER by 1\leqno{\fopen\the\FNUMBER\fclose}}
\def\RightFormulaLabel#1{\global\advance\FNUMBER by 1%
\eqno{\fopen\the\FNUMBER\fclose}\global\edef#1{\fopen\the\FNUMBER\fclose}}
\def\LeftFormulaLabel#1{\global\advance\FNUMBER by 1%
\leqno{\fopen\the\FNUMBER\fclose}\global\edef#1{\fopen\the\FNUMBER\fclose}}
%
%
%
\def\HeadNumber{\ifnum\CHAPTER>0 \the\CHAPTER.\fi%
\ifnum\SECTION>0 \the\SECTION.\ifnum\SUBSECTION>0 \the\SUBSECTION.\fi\fi}
\def\ComposedRightFormulaNumber{\global\advance\FNUMBER by 1%
\eqno{\fopen\HeadNumber\the\FNUMBER\fclose}}
\def\ComposedLeftFormulaNumber{\global\advance\FNUMBER by 1%
\leqno{\fopen\HeadNumber\the\FNUMBER\fclose}}
\def\ComposedRightFormulaLabel#1{\global\advance\FNUMBER by 1%
\eqno{\fopen\HeadNumber\the\FNUMBER\fclose}%
\global\edef#1{\fopen\HeadNumber\the\FNUMBER\fclose}}
\def\ComposedLeftFormulaLabel#1{\global\advance\FNUMBER by 1%
\leqno{\fopen\HeadNumber\the\FNUMBER\fclose}%
\global\edef#1{\fopen\the\FNUMBER\fclose}}
%
%
%
\def\TheoremNumber{\global\advance\FNUMBER by 1 \fopen\the\FNUMBER\fclose}
\def\TheoremLabel#1{\global\advance\FNUMBER by 1\fopen\the\FNUMBER\fclose%
\global\edef#1{\fopen\the\FNUMBER\fclose}}
%
%
%
\def\ComposedTheoremNumber{\global\advance\FNUMBER by 1 \fopen\HeadNumber\the\FNUMBER\fclose}
\def\ComposedTheoremLabel#1{\global\advance\FNUMBER by 1\fopen\HeadNumber\the\FNUMBER\fclose%
\global\edef#1{\fopen\HeadNumber\the\FNUMBER\fclose}}
%
%
%
\def\fopen{(}\def\fclose{)}                 
\def\fn{\ComposedRightFormulaNumber}        
\def\fl{\ComposedRightFormulaLabel}         
\def\tn{\ComposedTheoremNumber}             
\def\tl{\ComposedTheoremLabel}              

\def\Compare#1#2{\message{^^J Compara \noexpand #1:=#1[#2]^^J}}

%
%
%
%

\def\ni{\noindent}
\def\ss{\vskip 5pt}
\def\ms{\vskip 10pt}

\def\noex{\noexpand}

\def\refs{}
\def\empty{\#}
\def\BibNumber{}
\def\BibTitle{}

\newcount\BNUM
\BNUM=0

\def\bib#1#2{\gdef#1{\global\def\BibNumber{\empty}\global\def\BibTitle{#2}}}

\def\ref#1{#1
\if\BibNumber\empty \global\advance\BNUM 1
\message{reference[\BibNumber]}\message{}
\global\edef\refs{\refs \ss\ni[\the\BNUM]\ \BibTitle}
\global\edef#1{\noex\global\noex\edef\noex\BibNumber{[\the\BNUM]}
 \noex\global\noex\edef\noex\BibTitle{\BibTitle}}
{\bf [\the\BNUM]}
\else
{\bf \BibNumber}
\fi}

\def\Biblio{{\refs}}


\def\titolo{Remarks on N\"other charges and black holes entropy}
\def\autori{%
L.\ FATIBENE\note{E-mail: fatibene@dm.unito.it},
M.\ FERRARIS\note{E-mail: ferraris@dm.unito.it},
M.\ FRANCAVIGLIA\note{E-mail: francaviglia@dm.unito.it},
M.\ RAITERI\note{E-mail: raiteri@dm.unito.it}.}

\def\indirizzo{Dipartimento di Matematica, Universit\`a degli Studi di Torino,\par
\ni Via Carlo Alberto 10, 10123 Torino, Italy}

\def\Abstract{
We criticize and generalize some properties of N\"other charges presented in a paper
by V. Iyer and R. M. Wald and their application to entropy of black holes.
The first law of black holes thermodynamics is proven for any gauge-natural field
theory.
As an application charged Kerr-Newman solutions are considered.
As a further example we consider a $(1+2)$ black hole solution.}

\def\piedipagina{}


\def\Lor{\hbox{\rm Lor}}

\def\Div{\hbox{\rm Div}}
\def\SU{\hbox{\rm SU}}
\def\ds{\hbox{\bf ds}}
\def\d{\hbox{\rm d}}
\def\dt{\hbox{\rm dt}}
\def\dr{\hbox{\rm dr}}

\def\dphi{\hbox{\rm d$\phi$}}
\def\Aut{\hbox{\rm Aut}}
\def\dim{\hbox{\rm dim}}
\def\Con{\hbox{\rm Con}}

\def\fLie{\Xi_{\hbox{\it \$}}}
\def\Lie{\hbox{\it \$}}
\def\calP{{\cal P}} 
\def\calC{{\cal C}} 
\def\calE{{\cal E}} 
\def\calL{{\cal L}} 
\def\calW{{\cal W}} 
\def\calJ{{\cal J}} 
\def\calU{{\cal U}} 
\def\calB{{\cal B}} 
\def\calQ{{\cal Q}} 
\def\calS{{\cal S}} 
\def\calM{{\cal M}} 
\def\bfB{\hbox{{\bf B}}} 
\def\bftildeB{\hbox{{\bf\~ B}}} 


\def\na{\nabla}
\def\la{\lambda}
\def\Si{\Sigma}
\def\si{\sigma}
\def\ka{\kappa}
\def\Om{\Omega}
\def\om{\omega}
\def\ep{\epsilon}
\def\al{\alpha}
\def\be{\beta}
\def\Ga{\Gamma}
\def\ga{\gamma}

\def\de{\delta}
\def\ze{\zeta}
\def\te{\theta}

\def\R{{\Bbb R}}


\def\R{I \kern-.36em R}
\def\E{I \kern-.36em E}
\def\F{I \kern-.36em F}
\def\Co{I \kern-.66em C}
\def\id{1 \kern-.36em I}              

\def\del{\partial}                   

\def\QDE{{\offinterlineskip\lower1pt\hbox{\kern2pt\vrule width0.8pt
\vbox to8pt{\hbox to6pt{\leaders\hrule height0.8pt\hfill}\vfill%
\hbox to6pt{\hrulefill}}\vrule\kern3pt}}}
\def\um{\mathop{=}\limits}            


\def\arr{\rightarrow }            
\def\then{\quad\Rightarrow\quad}      
\def\QDE{\hbox{\ }\vrule height4pt width4pt depth0pt}                                                              

\def\np{\vfill\eject}
\def\ni{\noindent}

\def\ss{\vskip 5pt}
\def\ms{\vskip 10pt}
\def\bs{\vskip 15pt}

\bib{\WaldA}{V. Iyer and R. Wald, Phys. Rev. D {\bf 50},  1994, 846}

\bib{\WaldB}{R.M.\ Wald, J.\ Math.\ Phys., {\bf 31}, 1993, 2378 }

\bib{\WaldC}{B.\ S.\ Kay, R.\ M.\ Wald, Physics Reports {\bf 207}, (2),
North-Holland, 1991, 49}


\bib{\CADM}{M.\ Ferraris and M.\ Francaviglia, Atti Sem. Mat. Univ. Modena, {\bf 37},
1989, 61} 

\bib{\CADMB}{M.\ Ferraris, M.\ Francaviglia and I.\ Sinicco, Il Nuovo Cimento, {\bf 107B},
(11), 1992, 1303}

\bib{\CADMC}{M.\ Ferraris and M.\ Francaviglia, Gen.\ Rel.\ Grav., {\bf 22}, (9), 1990}

\bib{\Lagrange}{M. Ferraris, M. Francaviglia,
in: {\it Mechanics, Analysis and Geometry: 200 Years after Lagrange},
Editor: M. Francaviglia, Elsevier Science Publishers B.V., 1991}

\bib{\Ferraris}{M.\ Ferraris, in: {\it Geometrical Methods in Physics}, Proceedings of
the {\it Conference on Differential Geometry and Its Applications}, Nov\'e M\v esto na Morav\v e,
Czechoslovakia, September 5-9, 1983}

\bib{\Cavalese}{M. Ferraris and M. Francaviglia, in: {\it 8th Italian
Conference on General Relativity and Gravitational Physics}, Cavalese (Trento), August 30 --
September 3, World Scientific, Singapore, 1988, 183 }

\bib{\Bosoni}{L. Fatibene, M. Ferraris, M. Francaviglia,
J. Math. Phys. {\bf 38} (8), 1997, 3953}

\bib{\Stress}{M. Ferraris, M. Francaviglia,
J.\ Math.\ Phys., {\bf 26} (6), 1985}

\bib{\antichi}{A. Trautman, Comm.\ Math.\ Phys., {\bf 6} 1967, 248--261;
I. Kol\'a\u r, J.\ Geom.\ Phys., {\bf 1}, 1984, 127}

\bib{\BosoB}{L. Fatibene, M. Ferraris, M. Francaviglia,
{\it Do Bosons feel spin frames?}, Quaderni del Dip. Mat., {\bf 5}, 1998}

\bib{\Bologna}{L.\ Fatibene and M.\ Francaviglia, in: {\it Seminari di Geometria 1996/97},
Univ.\ Bologna, 1998}

\bib{\Jadwisin}{L.\ Fatibene and M.\ Francaviglia, in: {\it Procs.\ Gauge Theories of
Gravitation}, Jadwisin Sept.\ 1997, Acta Physica
Polonica B{\bf 29} (4), 1998, 915}

\bib{\PhD}{L.\ Fatibene, {\it Gauge-Natural Formalism for Classical Field Theories}, Ph.D. Thesis (to
appear)}

\bib{\Mottini}{M.\ Ferraris, M.\ Francaviglia, M.\ Mottini, Rend.\ Matematica, VII, {\bf 14}, Roma
1994, 457}

\bib{\Robutti}{M.\ Ferraris, M.\ Francaviglia and O.\ Robutti, in:{\it G\'eom\'etrie et Physique},
Proceedings of the {\it Journ\'ees Relativistes 1985} (Marseille, 1985), 112 -- 125; Y.\
Choquet-Bruhat, B.\ Coll, R.\ Kerner, A.\ Lichnerowicz eds. Hermann, Paris, 1987}

\bib{\Reina}{M.\ Ferraris, M.\ Francaviglia and C.\ Reina, J.\ Math.\ Phys.\ {\bf 24}, (1983), 120;
M.\ Ferraris, M.\ Francaviglia and C.\ Reina, Ann.\ Inst.\ H.\ Poincar\'e {\bf 38}, 1983,
371}

\bib{\MagnanoA}{M.\ Ferraris, M.\ Francaviglia, G.\ Magnano,
Gen.\ Rel.\ Grav., {\bf 19}, 1987, 465}

\bib{\MagnanoB}{G.\ Magnano, M.\ Ferraris, M.\ Francaviglia,
Class.\ Quantum Grav., {\bf 7}, 1990, 261}

\bib{\MagnanoC}{G.\ Magnano, M.\ Ferraris, M.\ Francaviglia,
J.\ Math.\ Phys.\ {\bf 31} (2), 1990, 378}


\bib{\Kerr}{B.\ Carter, Phys.\ Rev., {\bf 174}, 1968, 1559}

\bib{\RevA}{A. Trautman, in: {\it Gravitation: An Introduction to Current Research}, L. Witten ed.
Weley, New York, 1962, 168}

\bib{\RevB}{J. N. Goldberg, in: {\it General Relativity and Gravitation. One hunderd years after
the Birth of Albert Einstein}, {\bf 1}, A. Held ed. Plenum Press, New York, 1980, 469 }

\bib{\Nother}{E.\ N\"other, K\"onigl. Gesel. Wiss., G\"ottingen Nachr., Math. Phys. Kl.,
1918, 235}

\bib{\Rosen}{N.\ Rosen, Phys. Rev. 57, 1940,147 }

\bib{\Komar}{A.\ Komar, Phys. Rev. {\bf 113}, 1959, 934 }

\bib{\ADM}{R.\ Arnowitt, S.\ Deser and C.\ W.\ Misner, in: {\it Gravitation: An Introduction
to Current Research}, L. Witten ed. Weley, New York, 1962, 227}

\bib{\RT}{T.\ Regge and C.\ Teitelboim,  Ann.\ Phys., {\bf 88}, New York, 1974, 286}

\bib{\Banados}{M.\ Ba\~nados, C.\ Teitelboim and J.\ Zanelli,  Phys.\ Rev.\ Lett., {\bf 69},
1992, 1849}

\bib{\Mann}{J.D.\ Brown, J.\ Creighton and R.B.\ Mann,
Phys.\ Rev.\ D{\bf 50},  1974, 6394; S.\ Carlip, J.\ Gegenberg and R.B.\ Mann,  Phys.\ Rev.\ D{\bf
51}, 12, 1995, 6854}


\bib{\KrupkaA}{D. Krupka,
Folia Fac. Sci. Nat. UJEP Brunensis (Physica)
{\bf 14}, 1973, 1}
 
\bib{\KrupkaB}{D. Krupka,
{\it Semester on Diff. Geom.}, Sept.--Dec. 1979;
Internat. J. Theoret. Phys. {\bf 17}, 1978, 359;
Internat. J. Theoret. Phys. {\bf 15}, 1976, 949
}

\bib{\Kolar}{I.\ Kol{\'a}{\v r}, P.\ W.\ Michor, J.\ Slov{\'a}k, 
{\it Natural Operations in Differential Geometry}, 
Springer--Verlag, New York, 1993}

\bib{\KolarB}{M.\ Hor\'ak, I.\ Kol\'a\v r, Czech. Math. J., {\bf 33} (108), 1983, 467}


\bib{\Garcia}{P.\ L.\ Garc\'\i a, Symposia Math., {\bf 14}, Academic Press, London, 1976, 219}

\bib{\GarciaB}{P.\ L.\ Garc\'\i a, J. Mu\~noz, in: {\it Proceedings of the IUTAM-ISIMM Symposium on
Modern Developments in Analytical Mechanics}, Torino July 7--11, 1982; S.\ Benenti, M.\
Francaviglia and A. Lichnerowicz eds., Tecnoprint, Bologna, 1983, 127}

\bib{\boyer}{R.\ H.\ Boyer, Proc.\ Roy.\ Soc.\ A.\ {\bf 311}, 1969, 245}

\bib{\Gotay}{M.J.\ Gotay, J.\ Isemberg, J.E.\ Marsden, R.\ Montgomery,
{\it Momentum Maps and Classical Relativistic Fields. The Lagrangian and Hamiltonian
Structure of Classical Field Theories with Constrains.}}

\bib{\Anderson}{I.M.\ Aderson, in: {\it Mathematical Aspects of Classical
Field Theories}, ed.\ by M.\ Gotay et al., Cont.\ Math., {\bf 132}, 1992, 51}

\bib{\Katz}{J.\ Katz, Class.\ Quantum Grav., {\bf 2}, 1985, 423}

\bib{\Carter}{B.\ Carter, in: {\it Black Holes, Les Astres Occlus}, ed. C.\ DeWitt and  B.S.\
DeWitt, Gordon and Breach Science Publisher, New York, 1973, 57}

\bib{\Maeda}{J.-I.\ Koga, K.-I.\ Maeda, {\it
Equivalence of black hole thermodynamics between
a generalized theory of gravity and the Einsten theory},
e-print:gr-qc/9803086}

\bib{\String}{S. Ferrara and R. Kallosh, Phys. Rev. {\bf D 54}, 1996, 1514;
S. Ferrara and R. Kallosh, Phys. Rev. {\bf D 54}, 1996, 1525}

\vglue 103.3pt
\raggedbottom

\MainHead

\np
\pagenumbers


\NewSection{Introduction}

An expression for entropy of black holes has recently been proposed in a paper by Iyer
and Wald (see \ref{\WaldA}\ and references quoted therein).
This paper relates variations of entropy of stationary black holes (then generalized to the
non-stationary case) to variations of a suitable conserved quantity.

This idea relies on a well known and by now classical analogy between thermodynamic
laws and the evolution law for the horizon area of black holes.
However, we were intrigued by the relation established in \ref{\WaldA} between entropy
and N\"other charges expecially in view of a number of unclear questions, which we
hope to clarify here. In an appendix we summarize the main results presented in
\ref{\WaldA}, suitably restated with the notation we shall use in the next Sections (see
also \ref{\CADM}\ and \ref{\Lagrange}).
\def\CorrectEntropy{(10.14)} 
\def\EntropyDef{(10.7)} 
\def\NewPot{(10.12)} 
\def\KomSup{(10.3)} 
\def\IntMT{(10.10)} 
\def\MainTheorem{(10.9)} 

First of all, we claim that problems about the definition of integrals
(i.e. on the choice of representatives for superpotentials)
have been only partially faced in \ref{\WaldA}, where the solutions proposed are
insufficiently motivated. We also remark that the results of \ref{\WaldA}\ were achieved
in a somewhat ``twisted'' way. In fact, field equations, as well as the expression for
their solutions, were there used also when there was no need of doing so.
Moreover, a particular class of Lagrangians has been chosen with the aim of applying
the  framework to General Relativity;
the representatives of this class of Lagrangians depend on fields
and their derivatives through involved expressions which make results difficult
to be achieved.
Finally, most of the properties of N\"other charges used more or less explicitely in
\ref{\WaldA} have been in fact known for long time in various frameworks, some of which
much simpler and more general than the framework introduced in \ref{\WaldA}\  (see, for
example, \ref{\Lagrange}, \ref{\Stress},  \ref{\RevA}, \ref{\RevB}, \ref{\Cavalese}, 
\ref{\Bosoni}\ and references quoted therein).

For these reasons, in addition to the interest for applications not only in Relativity
but also in much more general frameworks of current physical interest, we believe that
 the property  $\MainTheorem$ in the appendix is worth being proved in a much
more general framework using all  the simple but powerful modern
mathematical tools which are naturally involved in it
 (and this will also enlight all problems in the definition of integrals). In particular
our attitude towards field theories brings us to use the geometric language of fiber
bundles in which variational calculus is naturally formulated.

This framework allows us in fact to clearly distinguish between quantities and
properties which hold only along classical solutions (i.e.\ {\it on-shell}) from
quantities and properties which hold on general configurations (i.e.\ {\it off-shell}).
The reasons for pursuing this distinction are not only {\sl mathematical},
but they are {\sl physical} reasons mainly.
In fact it is generally accepted that in quantum field theory (QFT)
physical contributions may come from configurations which are not classical solutions.
Thus, at least at QFT level, two quantities which differ off-shell may have very different physical
meaning even though they are classically identical (i.e.\ they coincide on-shell).

Moreover, the needless use of fields equations may break down properties which hold off-shell
(e.g.\ see below for {\it strong conservation laws}) or hide properties which
instead hold only on-shell. As an example, we can remark that global superpotentials,
which always exist in {\it natural theories} as General Relativity, show that
conserved currents are not only closed forms on-shell, but they are also {\it exact}
forms on-shell (regardless of the topology of space-time $M$).
Well, this is a relatively simple result to be proved working with off-shell N\"other currents
(also known as {\it momentum maps}; see \ref{\Gotay}),
but it is relatively difficult to produce a potential if one works with differential forms on
space-time instead of using forms on (suitable prolongations of) the configuration bundle.
(Of course expressions for on-shell currents do not change, thus they were and thus remain exact
forms; however, it often happens that just local potentials are built or, even worse,
global potentials are produced but they are claimed to exist just locally.) 

The requirement that the superpotentials $\calU(L,\xi,\si)$ are the pull-backs of the
{\it same} bundle theoretic quantity $\calU(L,\xi)$, which does not depend on the
section, also enlights a number of definition problems.
In fact $\calU(L,\xi)$ is global, so that integrals on boundaries are well defined;
canonical ways to construct $\calU(L,\xi)$ are provided in the correct framework, so
that also integrals
on closed domains are defined; finally our approach ensures regularity conditions on the
way $\calU(L,\xi,\si)$ depends on the section $\si$, which make variations of entropy
meaningful.

\NewSection{Natural and gauge-natural theories}

Let then $\calC=(C,M,\pi,F)$ be the {\it configuration bundle} of a field theory.
$\calC$ is a bundle over a (space-time) manifold $M$ which is assumed to be orientable,
paracompact and of dimension $n$. A {Lagrangian of order $k$} is a morphism $$
L:J^k\calC\arr A_n(M)
\fn
$$
from the {\it Lagrangian phase bundle} $J^k\calC$ (namely, the $k$-order jet prolongation of the
configuration bundle, i.e.\ the bundle where fields live together with their partial
derivatives up to order $k$ included) into the bundle $A_n(M)$ of $n$-forms over $M$.

One can define the bundle morphism
$\de L:J^k\calC\arr V^\ast(J^k\calC)\otimes A_n(M)$,
where $V^\ast(\cdot)$ denotes the dual of the vertical bundle, where {\it vertical} refers to
the relevant bundle projection (i.e.\ it denotes the kernel of the relevant tangent map, or
equivalently it refers to {\sl bundle-vectors which are tangent to fibers}).

The (global) morphism $\de L$ is defined by
$$
<\de L\circ j^k\si\>\vert\> j^kX>\>={d\over dt}\Big(L\circ J^k\hat\Psi_t\circ j^k\si\Big)\>
\Big\vert_{t=0}
\fn
$$
where $X$ is any vertical vector field on $\calC$, $\hat\Psi_t$ is its flow and
we denoted by $<\cdot\>\vert\>\cdot>$ the standard pairing between the dual vertical bundle
$V^\ast(\calC)$ (or some prolongation $V^\ast(J^k\calC)$ of it) and
the vertical bundle $V(\calC)$ (or $V(J^k\calC)$, respectively). 

It has been shown (see \ref{\Lagrange}\ and \ref{\antichi}) that for each Lagrangian $L$
the Calculus of Variations induces a unique
(global) morphism, called {\it the Euler-Lagrange morphism}
$$
\E(L)=J^{2k}\calC\arr V^\ast(\calC)\otimes A_n(M)
\fn
$$
together with a family of (global) morphisms (which depend on the Lagrangian and possibly on a
connection $\ga$ on $M$) called {\it Poincar\'e-Cartan morphisms}
$$
\F(L,\ga)=J^{2k-1}\calC\arr V^\ast(J^{k-1}\calC)\otimes A_{n-1}(M)
\fn
$$

The Euler-Lagrange morphism and the Poincar\'e-Cartan morphisms are in fact defined so that
the so-called {\it first-variation formula} holds for any vertical vector field $X$
on $\calC$:
$$
<\de L\>\vert\> j^k X>\>=\><\E(L)\>\vert\> X>+\>\Div<\F(L,\ga)\>\vert\> j^{k-1}X>
\fl{\FVF}
$$
where the {\it
formal divergence operator on forms} is defined by
$$
\Div(f)\circ j^{k+1}\si= \d(f\circ j^k\si),
\qquad\qquad f:J^k\calC\arr A(M)
\fn
$$
$\d(\cdot)$ being the exterior differential operator on forms and $A(M)\equiv\oplus_k A_k(M)$
denotes the bundle of forms over $M$.

We remark that the first-variation formula $\FVF$ is written {on the bundle $\calC$}
and it encompasses the whole class of {\it first-variation formulae} that can be written
on the base $M$ by pull-back along sections of the
configuration bundle $\calC$ (not necessarily solutions of field equations, i.e.\ of
the Euler-Lagrange equations $\E(L)\circ j^{2k}\si=0$).
We will show that performing calculations in this {\it off-shell fashion} turns out to
be important and certainly it clarifies relations among intrinsic quantities that can
be hidden by direct evaluation along solutions.

The Poincar\'e-Cartan morphisms are uniquely defined for $k=0,1$; for $k=2$ (which is
the case of interest for all metric theories of gravitation) they are not unique but
still there is a canonical choice which is independent on the connection $\ga$. For
$k\geq 3$ also the possibility of this canonical choice is lost and one has to work
with the whole class of Poincar\'e-Cartan morphisms (see \ref{\Garcia}, \ref{\GarciaB},
\ref{\Ferraris}\ and \ref{\KolarB}). Fortunately, in all physically relevant cases
we have $k\leq 2$ and therefore we shall speak of {\it the} Poincar\'e--Cartan
morphism (and form). 

If we consider a projectable vector field $\hat\Xi$ over $\calC$, i.e.\ a vector field
such that its projection $\xi=\xi(\hat\Xi)=\pi_\ast(\hat\Xi)$ is a well defined vector
field over $M$, we say that $\hat\Xi$ is an {\it infinitesimal symmetry of $L$} iff
$$
\Lie_\xi(L\circ j^k\si)=\><\de L\circ j^k\si\>\vert\> j^k\Lie_{\hat\Xi}\si>
\fn
$$
where $j^k\si$ denotes the $k$-order jet prolongation of any section $\si$ and the {\it
Lie-derivative of a section $\si$ with respect to $\hat\Xi$} is defined by the
following prescription:
$$
\Lie_{\hat\Xi}\si=T\si(\xi)-\hat\Xi\circ\si
\fn
$$

If the configuration bundle $\calC$ is a {\it natural bundle}, i.e. it is associated to
the $s$-frame bundle $L^s(M)$ for some $s\geq1$ (or, equivalently, its transition
functions depend functionally on local changes of coordinates in the base $M$ together
with their partial derivatives up to order at most $s$), we can restrict our attention
to projectable vector fields $\hat\xi$ on $\calC$ which are natural lifts of vector
fields $\xi$ on the base $M$. If all such vector fields $\hat\xi$ are infinitesimal
symmetries for the Lagrangian, we say that $L$ is a {\it natural Lagrangian} (see
\ref{\KrupkaA}, \ref{\KrupkaB}\ and references quoted therein). 

A field theory is said to be a {\it natural theory} if and only if both its configuration bundle
and its Lagrangian are natural.
An important example of natural theories is provided by the so-called {\it metric
theories of gravitation}, where the configuration bundle is chosen to be the bundle
$\calC=\Lor(M)$ of Lorentzian metrics on $M$, which is a natural bundle associated to
$L(M)$ ($s=1$); the prototype of metric theories is of course General Relativity.
The covariance principle of General Relativity just asserts that the Lagrangian is
natural and
the Hilbert-Einstein Lagrangian is a particular second-order natural Lagrangian
$(k=2)$.

However, not all physically relevant field theories are natural theories.
As an example, in Yang-Mills theories the gauge potentials are described by principal connections
on a $SU(m)$-principal bundle $\calP=(P,M,\pi,\SU(m))$. In general the bundle
$\Con(\calP)=J^1P/\SU(m)$ of principal connections on $\calP$ is not a natural bundle,
so that Yang-Mills theories cannot be natural. Moreover, in Yang-Mills theories pure
gauge symmetries are not the lift of diffeomorphisms on the base $M$ (they are all
vertical and thence project on the identity).

Let now $G$ be any Lie group and $\calP=(P,M,p,G)$ be a $G$-principal bundle, called the {\it
structure bundle}; we say that a bundle $\calC$ is a {\it gauge-natural bundle of order $(r,s)$
associated to $\calP$} (see \ref{\Kolar}, \ref{\Bologna}\ and references quoted therein) if it is
associated to the principal bundle $J^rP\times_M L^s(M)$ (with $s\geq r$),
where $\times_M$ denotes the fiber product over $M$.
A canonical action of the automorphisms of the structure bundle is defined on gauge-natural bundles
(and it completely characterizes them).
By means of this action, one can uniquely define for any infinitesimal generator $\Xi$
of automorphisms on $\calP$ a projectable vector field $\hat\Xi$ on the gauge-natural
bundle $\calC$ associated to the structure bundle.
We then say that the Lagrangian $L$ is {\it gauge-natural} if all such vector fields $\hat\Xi$
are infinitesimal symmetries of $L$.

We say that a field theory is a {\it gauge-natural theory} if both its configuration bundle
and its Lagrangian are gauge-natural.
Gauge-natural theories encompass, in a single and coherent framework, {\it all}
fields theories which are of interest in modern theoretical physics, since they
include in a unifying scheme all natural theories, gauge theories, Bosonic and
Fermionic matter as well as their mutual interactions. In order to achieve a canonical
treatment of conserved quantities (as shown in \ref{\Bologna}, \ref{\Jadwisin},
\ref{\PhD}\ and briefly recalled below) we also require that two principal connections
$(\ga,\om)$ on $L(M)$ and $\calP$, respectively, can be built out of the dynamical
fields. This latter requirement amounts to build two morphisms (which, by an abuse of
language, will still be denoted by $(\ga,\om)$) from (some jet prolongation of) the
configuration bundle $\calC$ to the bundles of linear connections $\Con({L(M)})$ and
gauge connections $\Con(\calP)$, respectively. Using these two morphisms, whenever we
fix a section $\si$ of $\calC$, we induce two connections on $L(M)$ and $\calP$, which
are thence called {\it dynamical connections}.
We stress that in higher order gauge-natural theories one can always
choose the representative of the Poincar\'e-Cartan morphism induced by the dynamical connection
on $M$, so that no ambiguity occurs.

As we said, most of the physically relevant field theories admit a formulation in the
gauge-natural framework. 
Clearly, natural theories are particular gauge-natural theories when $\calP$ is a $G$-principal
bundle with respect to the trivial group $G=\{e\}$ and the configuration bundle is a gauge-natural
bundle of order $(0,s)$ associated to $\calP$.
Analogously, pure gauge theories are gauge-natural provided one suitably reformulates them on
a gauge-natural bundle associated to
$J^1P\times_M L(M)$ (see \ref{\BosoB}).

\NewSection{Conserved quantities in gauge-natural theories}

Both in natural and gauge-natural theories conserved quantities are canonically
defined, because symmetries are encoded in the definition of the theory.
The aim of this Section is to extend to gauge-natural theories some properties of N\"other charges
that are known to hold in natural theories (see \ref{\WaldA}, \ref{\Lagrange}, \ref{\Komar},
\ref{\Nother}, \ref{\Rosen}\ and \ref{\Robutti}).

Let us consider the infinitesimal generator $\Xi$ of a 1-parameter subgroup of
automorphisms $\Phi_t$ of the structure bundle $\calP$.
As we said above, $\Xi$ induces a projectable vector field $\hat\Xi$ on the configuration bundle
$\calC$; $\hat\Xi$ is the infinitesimal generator of the flow $\hat\Phi_t$ induced on $\calC$ by the
flow $\Phi_t$ by means of the action of the group $\Aut(\calP)$ of automorphisms of $\calP$ on the
configuration bundle $\calC$ itself.

Since in gauge-natural theories fields are not natural objects, Lie-derivatives of fields with
respect to generic vector fields on $M$ are meaningless; however we can define
Lie-derivatives of fields with respect to infinitesimal generators of automorphisms on
the structure bundle $\calP$ as follows: $$
\Lie_\Xi\si:=\Lie_{\hat\Xi}\si=T\si(\xi)-\hat\Xi\circ\si,
\qquad\qquad \xi=p_\ast(\Xi)=\pi_\ast(\hat\Xi)
\fl{\LieSection}
$$
The Lagrangian is by definition $\Aut(\calP)$-covariant, i.e.\ it satisfies
$$
\Lie_\xi(L\circ j^k\si)=\><\de L\circ j^k\si\>\vert\> j^k\Lie_\Xi\si>
\fl{\CC}
$$

The first-variation formula $\FVF$ and the gauge covariance condition $\CC$ alone
provide a complete treatment of conserved quantities in higher order field theories.
In fact, these two formulae provide an intrinsic version of N\"other theorem by simply noticing
that the Lie derivative of a section $\si$ is vertical. Hence one can apply the
first-variation formula
$$
\eqalign{
\Lie_\xi&(L\circ j^k\si)=\><\de L\circ j^k\si\>\vert\> j^k\Lie_\Xi\si>\>=\cr
&=\><\E(L)\circ j^{2k}\si\>\vert\> \Lie_\Xi\si>
+\>\Div<\F(L,\ga)\circ j^{2k-1}\si\>\vert\> j^{k-1}\Lie_\Xi\si>\cr
}
\fl{\NoT}
$$
which, by expanding the l.h.s.\ Lie-derivative, allows to define the following currents:
$$
\eqalign{
\calE(L,\Xi)&=\><\F(L,\ga)\>\vert\> j^{k-1}\fLie>-i_\xi\> L\cr
\calW(L,\Xi)&=-<\E(L)\>\vert\> \fLie>\cr
}
\fl{\Currents}
$$ 
Here $\fLie:J^1\calC\arr V(\calC)$ is the global bundle morphism intrinsecally defined by
$$
(j^1\si)^\ast \fLie= \Lie_\Xi\si
\fn$$
where $\Lie_\Xi\si$ is the Lie derivative of a section $\si$ defined by $\LieSection$.

Because of $\NoT$, these two currents obey a conservation law
$$
\Div\>\calE(L,\Xi)=\calW(L,\Xi)
\fl{\ConsCurrOffShell}
$$

Let us then denote by $\calW(L,\Xi,\si)$ and $\calE(L,\Xi,\si)$ the pull-back of
$\calW(L,\Xi)$ and $\calE(L,\Xi)$, respectively, along a section
$\si$ of $\calC$.
Since $\calW(L,\Xi,\si)$ vanishes because of $\Currents$ whenever $\si$ is a solution of field
equations $\E(L)\circ j^{2k}\si=0$,  $\calE(L,\Xi,\si)$ is conserved {\it on-shell}.
In other words, one can pull-back eq.\ \ConsCurrOffShell\ along solutions producing a whole
class of currents (one for each section) which are closed forms on $M$.
We remark that eq.\ \ConsCurrOffShell\ is again written on $\calC$ (and not on $M$),
a fact which will turn out to be important below when discussing superpotentials.

The currents we defined, namely $\calE(L,\Xi)$ and $\calW(L,\Xi)$, are linear with respect to the
(infinite) jet prolongation $j\>\Xi$ of the infinitesimal generator of automorphisms
$\Xi=\xi^\mu\del_\mu+\xi^A\rho_A$ (here $\rho_A$ is a local basis for right invariant vertical vector
fields on $\calP$).
As a consequence, if the configuration bundle is a gauge-natural bundle of order
$(r,s)$, Lie derivatives of sections can  be written as linear combinations of symmetrised
covariant derivatives with respect to the dynamical connections $(\ga,\om)$
$$
\xi^\mu,\ \na_{\rho_1}\xi^\mu,\ \dots,\ \na_{\rho_1\dots\rho_s}\xi^\mu,
\xi^A,\ \na_{\rho_1}\xi^A,\ \dots,\ \na_{\rho_1\dots\rho_r}\xi^A
\fn$$
and, consequently, currents can be written as
$$
\eqalign{
\calE&(L,\Xi)=\Big[
T^{\la}_\mu\xi^\mu+T^{\la\rho_1}_\mu\na_{\rho_1}\xi^\mu+\dots
+T^{\la\rho_1\dots\rho_{k+s-1}}_\mu\na_{\rho_1\dots\rho_{k+s-1}}\xi^\mu+\cr
&+T^{\la}_A\xi^A+T^{\la\rho_1}_A\na_{\rho_1}\xi^A+\dots
+T^{\la\rho_1\dots\rho_{k+r-1}}_A\na_{\rho_1\dots\rho_{k+r-1}}\xi^A
\Big]\ds_\la\cr}
\fn$$
$$
\eqalign{
\calW&(L,\Xi)=\Big[
W^{}_\mu\xi^\mu+W^{\rho_1}_\mu\na_{\rho_1}\xi^\mu+\dots
+W^{\rho_1\dots\rho_{s}}_\mu\na_{\rho_1\dots\rho_{s}}\xi^\mu+\cr
&+W^{}_A\xi^A+W^{\rho_1}_A\na_{\rho_1}\xi^A+\dots
+W^{\rho_1\dots\rho_{r}}_A\na_{\rho_1\dots\rho_{r}}\xi^A
\Big]\ds\cr
}
\fl{\WorkExpansion}$$
where $\ds=\d x^1\land\dots\land\d x^n$ is the local volume form on $M$, while
$\ds_\mu=i_{\del_\mu}\>\cdot\>\ds$ are the local generators of $(n-1)$-forms. 
The coefficients ($T^{\la}_\mu$, $T^{\la\rho_1}_\mu$, $\dots$,
$T^{\la\rho_1\dots\rho_{k+s-1}}_\mu$, $T^{\la}_A$, $T^{\la\rho_1}_A$, $\dots$,
$T^{\la\rho_1\dots\rho_{k+r-1}}_A$) as well as ($W^{}_\mu$, $W^{\rho_1}_\mu$, $\dots$,
$W^{\rho_1\dots\rho_{s}}_\mu$, $W^{}_A$, $W^{\rho_1}_A$, $\dots$,
$W^{\rho_1\dots\rho_{r}}_A$) are {\it tensor densities} with respect to automorphisms of
the structure bundle.

Whenever we have such a linear combination we can perform covariant integration by parts
to obtain for the same quantity an equivalent linear expansion whose coefficients are
all symmetric with respect to upper indices, while the integrated terms are all pushed into a formal
divergence (see \ref{\Robutti}).
For example, we can recast the current
$\calW(L,\Xi)$ as 
$$
\calW(L,\Xi)=\calB(L,\Xi)+\Div\>\tilde\calE(L,\Xi)
\fl{\REDef}$$
The current $\tilde\calE(L,\Xi)$ vanishes identically on shell because of
$\Currents$ and $\WorkExpansion$ and
$\calB(L,\Xi)$ turns out to be
identically vanishing along any section (i.e. off-shell).
This holds
because of $\Div^2=0$ and since $\Div\>\calB(L,\Xi)=0$ for arbitrary $\Xi$ implies $\calB(L,\Xi)=0$
when $\calB(L,\Xi)$ is symmetric with respect to upper indices. 
The quantity $\tilde\calE(L,\Xi)$ is usually called the {\it reduced current}, and the
identities 
$$
\calB(L,\Xi)=0
\fl{\GenBianchiID}$$
are called {\it generalized Bianchi identities}.

The same kind of covariant integration by parts (which by the way is a well defined,
canonical  and global operation in the bundle framework) is performed on the current
$\calE(L,\Xi)$ and we obtain
$$
\calE(L,\Xi)=\tilde\calE(L,\Xi)+\Div\>\calU(L,\Xi)
\fn
$$
where $\tilde\calE(L,\Xi)$ is {\it the reduced current} defined in $\REDef$.
Again we shall denote by $\tilde\calE(L,\Xi,\si)$ and $\calU(L,\Xi,\si)$ the pull-back
along a section $\si$ of $\tilde\calE(L,\Xi)$ and $\calU(L,\Xi)$, respectively;
$\calU(L,\Xi)$ is called {\it a superpotential}.
Obviously, because of $\Div^2=0$, superpotentials are not unique but rather defined modulo formal
divergences.
Once again both the reduced current and superpotentials are calculated on the bundle
$\calC$ and can be pulled-back on the base $M$ along any section. Reduced currents and
superpotentials are easily shown to be global, thanks to their tensorial character;
 if they are pulled-back along a solution $\si$ the
reduced current vanishes and the current $\calE(L,\Xi,\si)$ is not only a closed form, but it is
also exact on-shell (see \ref{\Robutti}).

In other words, while the current $\calE(L,\Xi)$ is conserved just along solutions,
the quantity $\calE(L,\Xi)-\tilde\calE(L,\Xi)=\Div\>\calU(L,\Xi)$ is conserved
along any section of $\calC$.
We express this different behaviour by saying that $\calE(L,\Xi)$ is {\it weakly
conserved} while $\Div\>\calU(L,\Xi)$ is said to be {\it strongly conserved}.

As it is well known, natural theories as well as gauge-natural theories always allow (global)
superpotentials (see \ref{\PhD}\ and \ref{\Robutti}).
Thus one can say that in this kind of theories conserved currents are exact forms on-shell
and a representative for their potentials is explicitly produced.
This is a further result which can be completely hidden by performing calculations
on the base $M$.

Let us finally consider a {\it region} $D\subset M$ of spacetime, i.e. a compact $(n-1)$-submanifold
with a boundary $\del D\subset D\subset M$ which is a compact $(n-2)$-submanifold;
let us define {\it the conserved quantity} along a section $\si$ as
$$
Q_D(L,\Xi,\si)=\int_D \calE(L,\Xi,\si)=\int_{ D}\tilde\calE(L,\Xi,\si)+
\int_{\del D} \calU(L,\Xi,\si)
\fl{\CQNother}
$$

If $\si$ is a solution of field equations, the reduced current $\tilde\calE(L,\Xi,\si)$
vanishes and $Q_D(L,\Xi,\si)$ is just given by the integral of the superpotential on the
boundary of the region $D$.
In natural theories, the vector field $\Xi$ is usually taken to be the natural lift $\hat\xi$
of a vector field $\xi$ over the base $M$. In this case one simply writes
$\calU(L,\xi)$ instead of $\calU(L,\hat\xi)$ and so on, for notational convenience.

\NewSection{Intrinsic ADM formalism}\edef\ADMSection{\number\SECTION}

One of the standard approaches to conserved quantities in an Hamiltonian setting for General
Relativity is due to Arnowitt, Deser and Misner (ADM).
This method (see \ref{\ADM}) is motivated by the need of
selecting a preferred conserved quantity (the mass) which in the Hamiltonian formalism determines
the evolution of fields. This is achieved by a $(n+1)$ decomposition of space-time; this
decomposition is not unique and it destroys covariance of the theory, which has to be
restored at the end.
A covariant ADM formalism has been later developed (see e.g.\ \ref{\CADM}\ and \ref{\CADMB})
which does not require explicitly a $(n+1)$ decomposition.

In this framework, let us consider a current $\calE(L,\Xi)$ and a vertical vector field $X$ on
$\calC$; we calculate the variation of the current along $X$ and find:
$$
\eqalign{
\de_X&\calE(L,\Xi)=\de_X(<\F(L,\ga)\>\vert\> j^{k-1}\fLie>-i_\xi\> L)=\cr
=&\>\de_X<\F(L,\ga)\>\vert\> j^{k-1}\fLie>-\>i_\xi\><\de L\>\vert\> j^k X>\>=\cr
=&\>\de_X<\F(L,\ga)\>\vert\> j^{k-1}\fLie> -\>i_\xi\><\E(L)\>\vert\> X>+\cr
&\quad-i_\xi\Div\><\F(L,\ga)\>\vert\> j^{k-1} X>=\cr
=&\>\de_X<\F(L,\ga)\>\vert\> j^{k-1}\fLie>-\>\Lie_\xi\><\F(L,\ga)\>\vert\> j^{k-1} X>+\cr
&\quad
-i_\xi\><\E(L)\>\vert\> X>+\>\Div\>i_\xi<\F(L,\ga)\>\vert\> j^{k-1} X>\cr
}
\fl{\VarA}
$$
Integrating along a solution $\si$ we have then:
$$
\eqalign{
&\de_XQ_D(L,\Xi,\si)=\int_D\de_X\calE(L,\Xi,\si)=\cr
&=\int_D \de_X<\F(L,\ga)\circ j^{2k-1}\si\>\vert\> j^{k-1}\Lie_\Xi\si>+\cr
&-\int_D \Lie_\xi<\F(L,\ga)\circ j^{2k-1}\si\>\vert\>j^{k-1} X>+\cr &
-\int_D i_\xi\><\E(L)\circ j^{2k}\si\>\vert\> X>
+\int_{\del D} i_\xi<\F(L,\ga)\circ j^{2k-1}\si\>\vert\> j^{k-1} X>\cr}
\fl{\VarQ}
$$
The third integral on the r.h.s.\ vanishes because $\si$ is a solution. Following \ref{\CADM}, let us
now suppose
that boundary conditions for $X$ can be choosen so that a global $(n-2)$-form over $M$
$\bfB(L,\xi,\si)={1\over 2}B^{\mu\nu}\ds_{\mu\nu}$ exists and the following holds:
$$
\de_X \bfB(L,\xi,\si)\Big\vert_{\del D}=i_\xi<\F(L,\ga)\circ j^{2k-1}\si\>\vert\> j^{k-1}
X>\Big\vert_{\del D} \fn
$$
where, as usual, $\xi$ is the projection of $\Xi$ onto $M$.

As shown in \ref{\CADM}\ and \ref{\CADMB}, this is certainly the case of General Relativity.
Then we can redefine the conserved quantity $\CQNother$ by subtracting a boundary term
$$
\eqalign{
\hat Q_D(L,\Xi,\si)=&\int_D\calE(L,\Xi,\si)-\int_{\del D}\bfB(L,\xi,\si)=\cr
=&\int_D\tilde\calE(L,\Xi,\si)+\int_{\del D}[\calU(L,\Xi,\si)-\bfB(L,\xi,\si)]\cr}
\fl{\QADM}
$$
If $\si$ is a solution then the reduced current vanishes and $\hat Q_D(L,\Xi,\si)$
is a purely {\it ``boundary''} quantity.

Since in asymptotically simple space-times the quantity $\bfB(L,\xi,\si)$ has to be integrated
on spatial infinity, it 
 has to be a global  and therefore covariant $(n-2)$-form (otherwise the integral itself
would depend on coordinates and may not be defined if $\bfB(L,\xi,\si)$ is not defined on the
whole of spatial infinity). Of course it may happen that, in some particular
situation or for some special solution, some weaker recipe works; but, if this is the case, then it
cannot be claimed that these extend to {\it general} recipes.
Furthermore by using the non-uniqueness of $\calU(L,\Xi)$ and $\bfB(L,\xi)$ one suitably
imposes further conditions able to ensure that
the modified conserved quantities $\QADM$ vanish
on some particular field configuration, so to fix the closed-form ambiguity mentioned above.

In General Relativity we have the Lagrangian $L_{_{\rm H}}=r\sqrt{g}\>\ds$ ($r$ is the scalar
curvature of the Levi-Civita connection $\ga$ of the dynamical metric field $g$).
The Poincar\'e-Cartan morphism is (recall that $k=2$)
$$
<\F(L,\ga)\>\vert\>j^{1} X>\>=f^{\la\mu}_{\rho\si}\na_\mu X^{\rho\si}\>,
\qquad
f^{\la\mu}_{\rho\si}=\sqrt{g}\>[g^{\la\mu}g_{\rho\si}-\de^{(\la}_\rho\de^{\mu)}_\si]
\fn
$$
where $X^{\rho\si}=\de g^{\rho\si}$ is a vertical vector of $\Lor(M)$.

If we choose the standard boundary conditions $X\vert_{\del D}=0$, we can set
$$
\bftildeB=-{\sqrt{g}} g^{\al\be} \xi^{[\la}u^{\mu]}_{\al\be}\>\ds_{\la\mu}\>,
\qquad u^{\mu}_{\al\be}=\ga^{\mu}_{\al\be}-\ga^{\ep}_{\ep(\al}\de^\mu_{\be)}
\fl{\NBBCorrection}
$$

Unfortunately this choice of $\bftildeB$ is not covariant, i.e.\ $\bftildeB$
it is not a $(n-2)$-form ($n=\dim(M)$).
However, if one chooses a background connection $\Ga$, a covariant boundary
term exists and it is defined as follows:
$$
\eqalign{
\bfB(L,\xi,\si)=&-{\sqrt{g}} g^{\al\be} \xi^{[\la}w^{\mu]}_{\al\be}\>\ds_{\la\mu}\cr
&\qquad w^{\mu}_{\al\be}=u^{\mu}_{\al\be}-U^{\mu}_{\al\be},\quad
U^{\mu}_{\al\be}=\Ga^{\mu}_{\al\be}-\Ga^{\ep}_{\ep(\al}\de^\mu_{\be)}\cr} 
\fl{\FirstOrderCorrection}$$

We stress here that some background connection $\Ga$ is {\it always} chosen in current literature,
though usually for asymptotically flat solutions this is not evident since one generally fixes it
to be the Levi-Civita connection of Minkowski metric, which in suitable coordinates is even
vanishing. In fact, a background field  {\it has} to be chosen to provide covariance. If a
background connection is not chosen, this simply means that one is in fact fixing some particular
class of coordinate systems {\it and} implicitely a connection which is zero in {\it those}
coordinates.

As discussed in \ref{\CADM}\ and \ref{\CADMB}, the
conserved quantity $\hat Q_D(L,\Xi,\si)$ defined by \QADM\  has to be interpreted as {\it the
relative conserved quantity with respect to the background connection $\Ga$}.
According to this interpretation, $\Ga$ is considered unaffected by deformations $X$
while it is dragged by symmetry generators $\Xi$.
Of course the background connection $\Ga$ is not actually a physical field 
but rather a sort of ``gauge fixing''; in other words,
 it has to be
regarded as a mere {\it parameter} (i.e.\ we are interested, for instance, in the mass of the field
$g$ alone, and not in the total mass of $g$ and $\Ga$ together).
The interested reader may find discussed examples in \ref{\CADM}, \ref{\CADMB}, \ref{\CADMC}, whereby
a comparison among various background  choices and different prescriptions is also provided.
Application to the conserved quantities in spherically symmetric spacetimes can be found in
\ref{\Mottini}.

If the spacetime is asymptotically flat (according to one of the standard definitions), one can
choose $\Ga$ to be the Levi-Civita connection of Minkowski space and in this case one can recover the
standard ADM formalism.

In any case, these techniques apply to a much more general situation than asymptotically flat spaces.
If one performs now a $(n+1)$ decomposition (see \ref{\CADM}\ and \ref{\CADMB} for further
details) in the general case, one finds the boundary corrections known as {\it
Regge-Teitelboim terms} (see \ref{\RT}) plus some further boundary corrections (which of
course vanish in the asymptotically flat case).

\NewSection{First order General Relativity}\CurrentSection
\global\edef\FGR{\Mock}

As we said in the previous Section, ADM conserved quantities are defined by choosing
a background connection $\Ga$.
The same ingredients are needed to correct the so-called anomalous factors in Komar superpotentials
(see \ref{\Katz}) and to give a covariant first-order formulation of General Relativity
(see \ref{\Lagrange}, \ref{\CADMC}\ and references quoted therein).
We shall show that this mechanism is equivalent to intrinsic ADM formalism
as described above and then to standard ADM formalism when it applies.

As we have already done in the previous Sections, let us denote systematically by capital
letters the background quantities and by lower case letters the corresponding dynamical quantities,
i.e.
$$
\matrix{
g_{\mu\nu}\hfill       &     -\hfill         &\hbox{metric }\hfill    \cr
\ga^\al_{\be\nu}\hfill  &\Ga^\al_{\be\nu}\hfill &\hbox{connections }\hfill \cr
r^\al_{\be\mu\nu}\hfill  &R^\al_{\be\mu\nu}\hfill &\hbox{Riemann tensors }\hfill \cr
r_{\mu\nu} \hfill &R_{\mu\nu}\hfill &\hbox{Ricci tensors}\hfill \cr
r \hfill &R=g^{\mu\nu}R_{\mu\nu}\hfill &\hbox{scalar curvatures}\hfill \cr
u^\mu_{\al\be}= \ga^{\mu}_{\al\be}-\ga^{\ep}_{\ep(\al}\de^\mu_{\be)}\quad\hfill &
U^\mu_{\al\be}= \Ga^{\mu}_{\al\be}-\Ga^{\ep}_{\ep(\al}\de^\mu_{\be)}\hfill &\cr
}
\fn
$$ 
and let us define the relative quantities by:
$$
q^\mu_{\al\be}=\ga^{\mu}_{\al\be}-\Ga^{\mu}_{\al\be}\>,
\qquad\qquad
w^\mu_{\al\be}=u^{\mu}_{\al\be}-U^{\mu}_{\al\be}
\fn
$$

Since the Hilbert-Einstein Lagrangian is of the second order, field equations should be expected to
be of fourth order, while they are instead of the second order as if the Lagrangian were of the
first order only.
As is well known, this depends on the fact that second order derivatives can
be hidden in a divergence which does not affect field equations:
$$
L_{_{\rm H}}=\calL\>\ds=r\sqrt{g}\>\ds= \Big[p^{\al\be}(\ga^\rho_{\al\si}\ga^\si_{\rho\be}-
\ga^\si_{\si\rho}\ga^\rho_{\al\be})+d_\si(p^{\al\be}u^\si_{\al\be})\Big]\>\ds
$$ 
where we have set $p^{\al\be}=\sqrt{g}g^{\al\be}=\del\calL/\del R_{\al\be}$.

Unfortunately, this has only a local meaning since this splitting
of $L_{_{\rm H}}$ is not covariant. Again, if we use a background connection $\Ga$ as shown in
\ref{\CADMC}
we can achieve a covariant splitting $L_{_{\rm H}}=L_{_{\rm 1}}+
\Div\>(p^{\al\be}w^\si_{\al\be}\>\ds_\si)$, where
$$
\eqalign{
L_{_{\rm 1}}=&\Big[p^{\al\be}(\ga^\rho_{\al\si}\ga^\si_{\rho\be}-\ga^\si_{\si\rho}\ga^\rho_{\al\be})
+\d_\si (p^{\al\be}U^\si_{\al\be})\Big]\>\ds=\cr
=& \Big[R\sqrt{g}+ p^{\al\be}(q^\rho_{\al\si}q^\si_{\rho\be}-q^\si_{\si\rho}q^\rho_{\al\be})\Big]
\>\ds\cr}
\fn
$$
The first expression shows that the background $\Ga$ has no dynamics (so there is no restriction
to be fulfilled when we fix it); the second expression shows that $L_{_{\rm 1}}$ is covariant
(the quantities $q^\rho_{\al\si}$ being tensors on $M$).

Now we can calculate conserved currents for both $L_{_{\rm H}}$ and $L_{_{\rm 1}}$, finding:
$$
\eqalign{
&\calE_{{\rm H}}=\calE(L_{{\rm H}},\xi)= [p^{\al\be} \Lie_\xi u^\la_{\al\be}
-L_{{\rm H}}\xi^\la]\>\ds_\la = \tilde\calE_{{\rm H}}+\Div\>\calU_{{\rm H}}\cr
&\calE_{{\rm 1}}=\calE(L_{{\rm 1}},\xi)= [-w^\la_{\al\be} \Lie_\xi p^{\al\be}
-L_{{\rm 1}}\xi^\la]\>\ds_\la  = \tilde\calE_{{\rm 1}}+\Div\>\calU_{{\rm 1}}
\cr
}
\fn
$$
where we have set
$$
\eqalign{
&\tilde\calE_{{\rm H}}=2\sqrt{g}\Big(R_{\mu\nu}-{1\over 2}Rg_{\mu\nu}\Big)
g^{\mu\la}\xi^\nu\>\ds_\la\>,\cr
&\tilde\calE_{{\rm 1}}=\tilde\calE_{{\rm H}}
+p^{\al\be}\Lie_\xi U^\la_{\al\be}\>\ds_\la\>,
\cr
}
\qquad
\eqalign{
&\calU_{{\rm H}}= \na_\al\xi^{[\la}p^{\si]\al}\>\ds_{\la\si}\cr
&\calU_{{\rm 1}}=\calU_{{\rm H}}+\xi^{[\la}w^{\si]}_{\al\be}p^{\al\be}\>\ds_{\la\si}
}
\fn
$$
The superpotential $\calU_{{\rm H}}$ is a generalization of the Komar
potential (see expression $\KomSup$ in the appendix)
which was originally written for timelike Killing vectors (see \ref{\Komar}).

The background fixing produces an additional boundary term in
conserved quantities, namely
$$
Q_1=Q_H +
\int_{\del D} \xi^{[\la}w^{\si]}_{\al\be}p^{\al\be}\>\ds_{\la\si}
\fl{\QONE}
$$
which is exactly the same contribution to conserved quantities $\QADM$ due to the boundary correction
$\bfB(L,\xi,\si)$ given in ADM framework (see eq.\ $\FirstOrderCorrection$).
Consequently, both methods correct the anomalous factor problem
endemic to the Komar potential.
It has in fact been explicitely shown (see \ref{\Cavalese}\ and \ref{\CADMC}) that they provide the
correct values for both angular momentum and mass for Schwarzschild, Kerr and Kerr-Newman solutions.

These two methods are by no means {\it general}. The first method (ADM) uses space infinity
and thence it does not apply to compact solutions , like e.g.\ some cosmological solutions;  
an explicit recipe to build boundary corrections is then provided just for standard General Relativity,
though the method may as well apply to generalized metric theories, {\it provided} one gives
a way of building the boundary correction {\it and} exhibits a correct interpretation.
The second method has definitively stuck to standard General Relativity.

\NewSection{Variation of conserved quantities}

Let us now consider a gauge-natural field theory together with an infinitesimal generator $\Xi$ 
of automorphisms of the structure bundle $\calP$;
let $X$ be a vertical vector field on the configuration bundle $\calC$.
Thence one can define the variation of the current $\calE(L,\Xi)$ along $X$ as given by
$\VarA$.

On the other hand, since $\calE(L,\Xi)$ allows a superpotential we have
$$
\de_X\calE(L,\Xi)=\de_X\tilde\calE(L,\Xi)+\Div\>\de_X\calU(L,\Xi)
\fl{\VarB}
$$
and comparing eqs.\ \VarA\ and \VarB\ we obtain
$$
\eqalign{
\Div(&\de_X\calU(L,\Xi)-i_\xi<\F(L,\ga)\>\vert\> j^{k-1} X>)=\cr
=&\de_X<\F(L,\ga)\>\vert\> j^{k-1}\fLie>-\>\Lie_\xi\><\F(L,\ga)\>\vert\> j^{k-1} X>+\cr
&\quad-\de_X\tilde\calE(L,\Xi)
-i_\xi\><\E(L)\>\vert\> X>\cr
}
\fl{\MainWald}
$$

Eq.\ \MainWald\ holds off-shell; if it is pulled-back along a solution $\si$ then the
r.h.s. term $i_\xi\><\E(L)\circ j^{2k}\si\>\vert\> X>$ vanishes. 
The following two lemmas also hold.

\bs\ni{\bf Lemma \tl{\LemmaA}:}
If $\Lie_{\Xi} \si=0$ then
$$
\de_X<\F(L,\ga)\circ j^{2k-1}\si\>\vert\> j^{k-1}\Lie_\Xi\si>
-\>\Lie_\xi\><\F(L,\ga)\circ j^{2k-1}\si\>\vert\> j^{k-1}X>\>=0.
$$
\bs\ni
Outline of proof:
%
Since both $\de_X$ and $\Lie_\xi$ are derivatives the following holds
$$
\eqalign{
\de_X<\F(L,\ga)\circ j^{2k-1}\si\>\vert\>& j^{k-1}\Lie_\Xi\si>+\cr
-&\Lie_\xi<\F(L,\ga)\circ j^{2k-1}\si\>\vert\> j^{k-1}X>\>=\cr
=<\de_X\F(L,\ga)\>\circ j^{2k-1}\si\vert\>& j^{k-1}\Lie_\Xi\si>+\cr
&+<\F(L,\ga)\circ j^{2k-1}\si\>\vert\> j^{k-1}\de_X\Lie_\Xi\si>+\cr
&-\><\Lie_{\Xi}\F(L,\ga)\circ j^{2k-1}\si\>\vert\> j^{k-1}X>+\cr
&-\><\F(L,\ga)\circ j^{2k-1}\si\>\vert\> j^{k-1}\Lie_{\Xi}X>\>\cr
}
\fn
$$
Because of $\de_X\Lie_\Xi\si=\Lie_{\Xi}X$ the second and the fourth term on the r.h.s.\ cancel
each other, while $<\de_X\F(L,\ga)\circ j^{2k-1}\si\>\vert\> j^{k-1}\Lie_\Xi\si>\>$ vanishes
by the hypothesis $\Lie_{\Xi} \si=0$ and
$<\Lie_{\Xi}\F(L,\ga)\circ j^{2k-1}\si\>\vert\> j^{k-1}X>\>=0$ since $L$ is a gauge-natural
Lagrangian.\hfill\QDE

\bs
Let us consider the morphism $\de\E(L): J^{2k}\calC\arr V^\ast(J^{2k}\calC)\otimes
V^\ast(\calC)\otimes A_n(M)$.
We say that $X$ is a solution of the {\it linearized field equations} if
$$
\de\E(L)\>(j^{2k} X,  Y)=0
\qquad\qquad
\hbox{\rm for any vertical field $Y$ on $\calC$}
\fl{\linearized}
$$

\bs\ni{\bf Lemma \tl{\LemmaB}:}
If $\si$ is a solution of field equations and $X$ is a solution of linearized field equations
then $\de_X\tilde\calE(L,\Xi,\si)=0$.
\bs\ni
Proof:
We have
$$
\eqalign{
\de_X &\calW(L,\Xi)\um^{\Currents}-\de_X<\E(L)\>\vert\>  \fLie>=\cr
&=
-<\de_X\E(L)\>\vert\> \fLie >-<\E(L)\>\vert\> \de_X\fLie >=\cr
&=-\de\E(L)\>(j^{2k} X,  \fLie )-<\E(L)\>\vert\>  \de_X\fLie >\cr
}
\fn
$$
The last term $<\E(L)\>\vert\>  \de_X\fLie >$ vanishes on-shell.

We recall that integrating by parts $\calW(L,\Xi)$ (see $\REDef$)  and
using Bianchi identities $\GenBianchiID$, 
one obtains a pure divergence, i.e.\  $\calW(L,\Xi)=\Div\> \tilde\calE(L,\Xi)$;
this easily proves that
$$
0=\de\E(L)\>(j^{2k} X, \de_X\Lie_{\Xi}\si)=\de_X \calW(L,\Xi,\si)=
\Div\> \de_X\tilde\calE(L,\Xi,\si)
\fn
$$
This implies that $\de_X\tilde\calE(L,\Xi,\si)=0$, again because of the symmetry of the
coefficients.
Thence also Lemma $\LemmaB$ is proved.\hfill$\QDE$

\bs
Applying  these two lemmas to $\MainWald$, the following theorem  is then  proved:

\bs\ni{\bf Theorem \tn:}
If $\si$ is a solution of field equations, $\Lie_{\Xi} \si=0$ and $X$ is a solution of
linearized field equations $\linearized$ then
$$
\Div(\de_X\calU(L,\Xi,\si)-i_\xi<\F(L,\ga)\circ j^{2k-1}\si\>\vert\> j^{k-1} X>)=0
\fn$$

\bs
This theorem generalizes to all gauge-natural
theories the result that Iyer and Wald proved in
\ref{\WaldA} for a particular class of natural theories and which has been used in \ref{\WaldA}\ and
\ref{\WaldC}\ to define black holes entropy and to {\it prove} the first principle of thermodynamics.

As noticed by Iyer and Wald in \ref{\WaldA}, the objects involved in the theory of conserved
quantities, e.g.\ the superpotential, may be ambiguous.
In fact, one can add a total divergence to the Lagrangian
without actually changing field equations but changing the superpotential itself. As we said in
Section $\FGR$, this is the key property to {\it cure} the anomalous factor problem.

Adding a total divergence  $\Div\> \te\>$ to the Lagrangian $L$, we induce in fact an extra term
$i_\xi\te$ in the superpotential $\calU(L,\Xi)$.
Nevertheless, in the framework developed in \ref{\WaldA} this latter term does not cause any
ambiguity because $\xi$ vanishes on the bifurcation surface $\Si$ (see eq.\ $\EntropyDef$ in the
appendix). We stress that, in our framework, the relevant quantity (see $\CorrectEntropy$ in the
appendix) is instead
$$
\de_X\calU(L,\Xi)-i_\xi<\F(L,\ga)\>\vert\> j^{k-1} X>
\fl{\LF}$$
which again is not affected by any ambiguity since the extra term $\de_X(i_\xi\te)$
in the variation of the superpotential is exactly cancelled by the contribution  $i_\xi\de_X(\te)$
due to the second term in $\LF$. 

\eject
\NewSection{Kerr-Newman black holes}

We now  apply the  result of the previous  Section to treat the cases of Kerr-Newman solutions.
In some way this is a toy example, since, even though electromagnetic
field is involved (and thence a gauge-natural theory should 
in principal be needed), the electromagnetic potential
$A_\mu$ is at the same time a (local) $1$-form on $M$ as well as a $U(1)$-connection (see e.g.\ \ref{\Reina} for
more details on the naturality of Maxwell theory).
Thus if one gives up the description of electric charges (and (pure) gauge transformations) the
Einstein-Maxwell system can be treated as a natural theory and can be thence
described by standard results for natural theories.
If one wants to analyze truly gauge-natural theories one should cope with 
$\SU(n)$ Yang-Mills theories, but in
these cases some problems about the interpretation of exact solutions arise; in particular there
may be problems about the {\it correct conserved quantities} to be produced.

Let us then consider a gravity-electromagnetic (GEM) system.
Its configuration bundle is
$$
\calC= \Lor(M)\times_M \Con(\calP)
\fn
$$
where $\calP$ is the trivial $U(1)$-bundle over the maximal extension of the Kerr-Newman spacetime
(see, e.g.\ \ref{\Kerr}\ and \ref{\Carter}).
Local fibered coordinates on $\calC$ are $(x^\mu,g_{\mu\nu}, A_\mu)$.

The (GEM) system is described by the following Lagrangian
$$
L=\calL_{_{\rm H}}\>\ds+\calL_{_{\rm EM}}\>\ds= {\sqrt{g}\over 16\pi}\Big[ r_{\mu\nu}g^{\mu\nu}\>-
 F_{\mu\nu}F_{\rho\si}g^{\mu\rho}g^{\nu\si}\Big]\>\ds
\fn
$$
where $F_{\mu\nu}=\del_\mu A_\nu-\del_\nu A_\mu$ is the curvature of the Maxwell connection
$A_\mu$.

The bundle $\calC$ is a gauge-natural bundle of order $(1,1)$ associated to the structure bundle
$\calP$ and $L$ is a gauge-natural Lagrangian.
The dynamical connections are the Levi-Civita connection of $g_{\mu\nu}$ and the Maxwell connection $A_\mu$.

Let us consider a Kerr-Newman solution of the (GEM) system.
Ingoing Kerr-Schild coordinates $(t,r,\theta,\phi)$ on $M$ can be choosen so that the solution 
reads as
$$
\eqalign{
&g= \eta+ \rho^{-2}(2mr-e^2)\Big[\dt +\dr-a\sin^2\theta\dphi\Big]^2\cr
&A=-er\rho^{-2}\Big[\dt+\dr-a\sin^2\theta\dphi\Big]\cr}
\fl{\KNSolution}
$$
where $\rho^2=r^2+a^2\cos^2\theta$, $m^2\ge e^2+a^2$, and we have set
$$
\eta= -\dt^2+\Big[\dr-a\sin^2\theta\>\dphi\Big]^2+\rho^2\Big[\d\theta^2+\sin^2\theta\>\dphi^2\Big]
\fn
$$
The metric in $\KNSolution$ is singular in $\rho^2=0$ and has two
horizons $r_{\pm}=m\pm\sqrt{m^2-a^2-e^2}$. The singularity in $\rho^2=0$ is a true one, as one
can see, e.g., by computing the invariant $R_{\mu\nu\rho\si}R^{\mu\nu\rho\si}$ (see \ref{\Kerr}).

We remark that, when $m=0$ and $e=0$, $g$ reduces to $\eta$ which is the flat Kerr-Schild metric.
The Levi-Civita connection of $\eta$ is choosen as a background $\Ga$.
As a background for the gauge-field we shall choose the vanishing gauge potential $A^{_{(B)}}_\mu=0$.

The vector field $\Xi=\xi^\mu\del_\mu+\ze\hat\rho$, where $\hat\rho$ is the right invariant generator
of the Lie algebra $i\R$ of $U(1)$ on the structure bundle $\calP$, induces an infinitesimal
symmetry on $\calC$. The induced superpotential is of the following form:
$$
\eqalign{
\calU(L,\Xi)=&\Big[\na_\be\xi^{[\al} p^{\si]\be}\Big]\>\ds_{\al\si}\cr
&-\Big[(4\pi)^{-1}  F^{\al\si}_{\>\cdot\>\>\cdot} A_\mu\xi^\mu\Big]\sqrt{g}\>\ds_{\al\si}\cr
&-\Big[(4\pi)^{-1} F^{\al\si}_{\>\cdot\>\>\cdot}\ze\Big]\sqrt{g}\>\ds_{\al\si}\cr
}\fl{\KNSupPot}
$$
The first contribution is the generalized Komar potential, the second comes from the
electromagnatic field and the third comes from gauge invariance.
The boudary correction $\FirstOrderCorrection$ is given by
$$
\bfB(L,\xi)=-\xi^{[\al}w^{\si]}_{\mu\nu} p^{\mu\nu}\>\ds_{\al\si}
\fn$$
and the corrected superpotential is defined as follows:
$$
\calU_1(L,\Xi)=\calU(L,\Xi)-\bfB(L,\xi) 
\fn$$
This is the correct superpotential since $\Lie_\Xi\Ga=0$ and $\Lie_\Xi A^{_{(B)}}=0$
(if this were not the case, the reduced current would not vanish on shell because we would be changing also
the background with respect to which we are computing {\it ``conserved quantities''}).

If we consider $\del_t$ (which is a Killing vector for both backgrounds since
$\Lie_\Xi U^\si_{\mu\nu}=0$ and $\Lie_\Xi A^{_{(B)}}_\mu=0$) integrating on spatial spheres
$S^2_r=\{r=\hbox{\rm constant}> r_+\}$
we have the mass
$$
\calE= \int_{S^2_r} \calU_1(L,\del_t,\si)= m+{e^2r\over 2(a^2+r^2)}
\fn
$$
As long as $r$ goes to infinity, we get $\calE=m$, as expected.

If we consider $-\del_\phi$ (which is again a Killing vector for both backgrounds)
integrating on spatial spheres $S^2_r$ we have the angular momentum
$$
\calJ= -\int_{S^2_r} \calU_1(L,\del_\phi,\si)= ma
\fn
$$

If we consider $\Xi=b\hat\rho$ with $b$ constant (which is again a Killing vector for the background
since $g^{\mu\nu}$ is not affected by pure gauge transformations and $A^{_{(B)}}$ is invariant
because $b$ is constant) integrating on spatial spheres $S^2_r$ we have the electric charge
$$
\calQ= \int_{S^2_r} \calU_1(L,b\hat\rho,\si)=-be
\fn
$$

Let us now consider the vector field
$\Xi=\del_t+\Om_{_{\rm H}}\del_\phi+ b\hat\rho$
(where $\Om_{_{\rm H}}=a/(a^2+r_+^2)$ and $b=em/(2mr_+-e^2)$).
It is a Killing vector for the backgrounds, it leaves the
solution invariant (i.e.\ $\Lie_\Xi\si=0$); thus, if $X$ is a solution of 
the linearized field equations
(i.e.\ it is {\it ``tangent''} to the  space of solutions) we have (see $\IntMT$)
$$
\eqalign{
{\ka\over 2\pi}\de_X\calS=&
\int_{S^2_r} [\de_X \calU(L,\Xi,\si)-i_\xi<\F(L,\ga)\>\vert\> j^1 X>]
=\cr
=&\int_{\infty} \de_X[ \calU(L,\Xi,\si)-\bfB(L,\xi,\si)]
=\cr
=&\de_X\calE-\Om_{_{\rm H}}\de_X\calJ- b\de_X\calQ= (1-a\Om) \de m- m\Om \de a-b \de e
}
\fl{\FPT} $$
Setting now for the surface gravity
$$
\ka={\sqrt{m^2-a^2-e^2}\over 2mr_+ -e^2}
\fn$$
and integrating $\FPT$ one gets
$$
\calS=2\pi m r_+
\fl{\Entropy}$$
Using then the expressions for $(r_+,\Omega,b,\kappa)$
one can expand variations with respect to $(\de m,\de a,\de e)$ and verify directly that
$\Entropy$ satisfies $\FPT$.
We remark that in these coordinates (as well as in any other {\it usual} coordinates
but the Kruskal-like ones) the contribution from the term $i_\xi<\F(L,\ga)\>\vert\> j^1X>$
never vanishes on any sphere for any $r$.
This is because $\Si$, though belonging to the maximal extension of the Kerr-Newman solution, does
not intersect the domain of the coordinates used.
For this reason we believe that the definition $\CorrectEntropy$ in appendix
for the variation of entropy turns out to
be much more useful than its {\it simplification} $\EntropyDef$
in the appendix.
Moreover further hypotheses (as well as more calculations) are required to perform such a
simplification.

\NewSection{Another example}

As observed in \ref{\Banados} Einstein equations with (negative) cosmological constant
allow a $(1+2)$-dimensional black hole solution.
Let us consider the Lagrangian
$$
L=\alpha\Big( r +{2\over l^2}\Big)\sqrt{g}\>\ds
\fn$$
Considering its variation one finds
$$
\eqalign{
&<\E(L)\>\vert\> X>\>= \alpha \sqrt{g}\Big[r_{\mu\nu} -\Big({1\over 2}R+{1\over
l^2}\Big)g_{\mu\nu}\Big] X^{\mu\nu}\>\ds\cr
&<\F(L,\ga)\>\vert\> j^1X>\>= \alpha\sqrt{g} \Big[g^{\la\si}g_{\al\be}
-\de^\la_{(\al}\de^\si_{\be)}\Big]\na_\si X^{\al\be}\>\ds_\la\cr
&\tilde\calE(L,\xi)=2\alpha\sqrt{g} \Big[r^\la_{\cdot\nu} -\Big({1\over 2}R+{1\over
l^2}\Big)\de^\la_{\cdot\nu}\Big]\xi^\nu\>\ds_\la\cr
&\calU(L,\xi)=\alpha\sqrt{g} \na^{[\be}\xi^{\al]}\>\ds_{\al\be}\cr\cr
}
\fn$$
This model allows the following black hole solution
$$
g=-N^2\>\d t^2+ N^{-2} \d r^2 +r^2(N_\phi \d t +\d\phi)^2
$$
where we have set
$$
\eqalign{
&N^2=-M+{r^2\over l^2}+ {J^2\over 4r^2}\>,\cr
&N_\phi=-{J\over 2r^2}\>,\cr
}
\qquad
\qquad
\eqalign{
&M={r^2_++r^2_- \over l^2}\cr
&J={2r_+r_-\over l}\cr
}
\fn$$

%

Once again, if one computes the entropy as shown in appendix using
$\CorrectEntropy$ (choosing any $\Si'$ with $r=\hbox{constant}$
in the coordinate used)
in place of $\EntropyDef$, one easily gets:
$$
\de\calS= 4\pi^2\de r_+ \then \calS=4\pi^2 r_+\quad\qquad\Big(\ka={ r^2_+-r^2_-\over r_+ l^2}\Big)
\fn$$
which is exactly the entropy expected (see \ref{\Mann}).

If the contribution of the term $i_\xi<\F(L,\ga)\>\vert\> j^1X>$
is considered alone it can be easily checked that
this contribution cannot be discarded on any $S_r$.

\NewSection{Conclusions}

We have generalized the definition of entropy given in \ref{\WaldA}.
Our formalism applies not only to a general-covariant metric theory,
but to any gauge-natural theory.

In particular this, together with the Legendre equivalence of $R$,
$R^2$ and $R_{\mu\nu}R^{\mu\nu}$ theories proved in \ref{\MagnanoA}, \ref{\MagnanoB} and
\ref{\MagnanoC}  explains some recent results on black holes entropy
in non linear theories of gravitation by Maeda et al. (\ref{\Maeda}),
which will form the subject of further investigation.
Moreover, our formalism can be fruitfully applied to black hole
entropy in string theory, which is a hot subject in current literature
(see, e.g.\ \ref{\String}). Also this will form the subject of further
investigation.

The definition for the variation of the entropy is
$$
\de_X\calS={1\over T}\int_{\Si'} [\de_X\calU(L,\Xi,\si)-i_\xi <\F(L,\ga)\circ j^{2k-1}\si\>\vert\>
j^1 X>] \fl{\Claim}$$   
where $\Xi=\del_t+\Om \del_\phi+b^A\rho_A$ is a Killing vector for the solution $\si$
and $(\Om,b^A)$ are chosen to provide the expected form of the first principle of thermodynamics:
$$
\de_X\calE=T\de_X\calS+\Om\de_X\calJ+b^A\de_X\calQ_A
\fl{\GenFTP}$$
The domain of integration is than {\it any} $(n-2)$-surface $\Si'$, such that
$\infty-\Si'$ is a homological border (e.g.\ it does not include singularities). 

This prescription for entropy may be regarded exactly as an algorithm to define a quantity $\calS$
such that $\GenFTP$ holds.

We remark that the coefficients $(T,\Om,b^A)$ has  to be provided by some other argument.
In some case, under some severe hypotheses, the fully general prescription
$\Claim$ reduces to the simplified prescription given in $\ref{\WaldA}$.

\NewSection{Appendix}

This appendix is devoted to summarize and recast the partial results
presented in  \ref{\WaldA} into the general theory we presented above.
All the results reproduce {\it exactly} those of \ref{\WaldA}
and are just restated in our notation;
we stress that some imprecise statements (especially on locality
of potentials and some of their ambiguity) are to be compared with
their correct counterparts in our framework.

As is well known, if one considers a Lagrangian $L:J^k\calB\arr A_n(M)$,
then its variation defines two morphisms $<\E(L) \>\vert\> X>$ and
$<\F(L,\ga) \>\vert\> j^{k-1}X>$ such that the first variation formula holds:
$$
<\de L \>\vert\> j^kX>=<\E(L) \>\vert\> X>+\>\Div<\F(L,\ga) \>\vert\> j^{k-1}X>
\fn$$
Then $\E\circ j^{2k}\si=0$ are field equations, while $<\F(L,\ga) \>\vert\> j^{k-1}X>$ is a
$(n-1)$-form depending on fields and (linearly) on their deformation.
Though the quantity $<\F(L,\ga) \>\vert\> j^{k-1}X>$ does not play any role in field equations, it is known
to be tightly related to conserved quantities.
In fact, N\"other theorem claims that if an infinitesimal Lagrangian symmetry
$\xi$ is considered,
then one can define a conserved current $\calE(L,\xi,\si)=(j^{2k}\si)^\ast[<\F(L,\ga) \>\vert\>
j^{k-1}\Lie_\xi\si>-i_\xi L]$, where $\Lie_\xi\si$ denotes Lie derivative of fields.
The current $\calE(L,\xi,\si)$ is a closed $(n-1)$-form on spacetime $M$ ($\dim(M)=n$).
The current $\calE(L,\xi,\si)$ depends on the Lagrangian, on the symmetry
$\xi$ considered and on a solution $\si$ of field equations.
Since the currents $\calE(L,\xi,\si)$ are closed $(n-1)$-forms,
they allow potentials which are $(n-2)$-forms $\calU(L,\xi,\si)$ on $M$ 
locally constructed from fields and $\xi$ (see \ref{\Anderson}, \ref{\WaldB}) such that
$$
\d\>\calU(L,\xi,\si)=\calE(L,\xi,\si)
\fl{\PotDef}$$ 

Conserved quantities $\calQ_{D}(L,\xi,\si)$ are defined integrating
the conserved current $\calE(L,\xi,\si)$ on a {\it regular domain}
$D\subset M$.
More generally one can consider the integrals of the potential
$\calU(L,\xi,\si)$ on a closed $(n-2)$-submanifold $C$, which may not be
the boundary of a {\it regular domain} $D\subset M$
(here {\it boundary} is used in the homological sense).

Of course, the potentials $\calU(L,\xi,\si)$ are not
uniquely defined, since also $\calU'(L,\xi,\si)=\calU(L,\xi,\si)+\al(\si)$ is
a potential provided $\d\al(\si)=0$.
[Authors' note: if physical sense has to be given to integrals
of the potential $\calU(L,\xi,\si)$ on a closed region $C$ which are
not boundaries, then a canonical way of choosing a representative
$\calU(L,\xi,\si)$ has to be provided, the integral depending
on the form $\al(\si)$.
Of course if we change representatives then the integral values change.]

Specializing to standard General Relativity, a superpotential exists having the form
$$
\calU_{_{\rm Kom}}(L,\xi,\si)=\sqrt{g}\>\na^\mu\xi^\nu \ds_{\nu\mu}
\fl{\NKomSup}$$\Compare{\NKomSup}{\KomSup}
It was given by Komar (see \ref{\Komar}) for a timelike Killing vector $\xi$
and then generalized to an arbitrary vector field.
Here $\ds_{\mu\nu}$ denotes the standard local basis for $(n-2)$-forms. 

Let $\si$ be an asymptotically flat solution of standard General Relativity,
$t$ a Killing vector which is an asymptotic time translation
and $\la$ a Killing vector which is an asymptotic spatial rotation;
if one considers the conserved quantities  associated to them integrating 
the Komar superpotential on
{\it space infinity} (which is not a boundary because of singularities),
$\calQ_\infty(L,\la,\si)$ produces the angular momentum
but $\calQ_\infty(L,t,\si)$ gives just one half of the expected mass. 
This is commonly known as the {\it anomalous factor problem} (see \ref{\Katz})
which can be {\it cured} considering a suitable boundary term $\bfB_{_{\rm ADM}}(L,\xi,\si)$
such that
$$
\de_X \bfB_{_{\rm ADM}}(L,\xi,\si)\Big\vert_\infty=
i_\xi <\F(L,\ga)\>\vert\> j^1X>\Big\vert_\infty
\fn$$
Then one can correct the definition of conserved quantities in the following way
$$
\eqalign{
& \calM^{_{\rm ADM}}=\int_\infty [\calU_{_{\rm Kom}}(L,t,\si) -\bfB_{_{\rm ADM}}(L,t,\si)]
\cr
&\calJ^{_{\rm ADM}}=-\int_\infty [\calU_{_{\rm Kom}}(L,\la,\si) -\bfB_{_{\rm ADM}}(L,\la,\si)] \cr
}
\fl{\ADMDefinition}
$$
which both give the expected values since the correction on angular momentum
vanishes.
The recipe to define $\bfB_{_{\rm ADM}}(L,\xi,\si)$ was given by Arnowitt, Deser and Misner (ADM)
using a $(n+1)$ decomposition and it is generally known as {\it ADM formalism}.

\ss
In \ref{\WaldA}\ a general covariant metric theory is considered;
it is described by a Lagrangian of the form
$$
L=\calL(g_{\mu\nu}, R_{\al\be\mu\nu}, \na_\la R_{\al\be\mu\nu},\dots,
\na_{(\la_1}\dots\na_{\la_m)}R_{\al\be\mu\nu}), \>\ds
\fn$$
That is simply a particular class of natural (and thence gauge-natural)
Lagrangians.

A stationary, asymptotically flat solution $\si$ is considered and
assumed to have a bifurcate Killing horizon (see \ref{\WaldC}, \ref{\boyer}). 
Let us consider a Killing vector $\xi=t+\Om_{_{\rm H}}\la$ which
vanishes on the bifurcation $(n-2)$-surface $\Si$ which is the
only {\it internal boundary}. 
To correct anomalous factor, a boundary term $\bftildeB$
analogous to $\NBBCorrection$ is considered.
[Authors' note: the boundary correction $\bftildeB$ in \ref{\WaldA}
is explicitly not required to be covariant so that it is definitely local
and coordinate dependent!
The {\it ``corrected''} conserved quantities are thence 
undefined, being the integrand {\it a priori} not even defined
on the domain of integration. Even assuming that $\bftildeB$
is defined on the whole of space infinity, still the integral is coordinate dependent,
against the very basic prescriptions of General Relativity.
Alternatively, this quantities are, at best, related to some preferred
(and undefined) class of coordinate systems.
Of course these problems are avoided when introducing the background connection
$\Ga$ as shown in Section $\ADMSection$. We stress that strictly speaking
$\Ga$ has not to be regard as a background field, but as a parameter;
thus it does not violate prescriptions of General Relativity.
As we already explained, the conserved quantities one defines using $\Ga$
have to be interpreted as {\it relative} conserved quantities
{\it with respect to $\Ga$}.]

Then the entropy of this stationary black hole is implicitly defined by
$$
{\ka\over 2\pi}\de_X\calS=\int_\Si \de_X \calU(L,\xi,\si) 
\fl{\NEntropyDef}
$$\Compare{\NEntropyDef}{\EntropyDef}
where $X=\de g$ is a {\it deformation}, i.e.\ a vertical vector field,
and where $\ka$ is the {\it surface gravity} defined by
$$
\ka^2=-{1\over 2} \na^a\xi^b\na_a\xi_b
\fn$$
and it can be shown to be constant on $\Si$.

Of course $\de\calS$ is not well defined since $\Si$ is not a boundary so
that one has to explain
{\it why} (or at least {\it how}) a particular representative for
$\calU(L,\xi,\si)$ is chosen.
[Authors' note: of course Komar superpotential {\it is} a possible choice
but it is a superpotential
for {\it standard} General Relativity with the {\it standard} Hilbert-Einstein Lagrangian;
here $\calU(L,\xi,\si)$ is a superpotential of the general covariant metric theory
taken under consideration. In \ref{\WaldA}\ an algorithm is provided for building such
a representative, but it is there also claimed that it is {\it one} of many possible algorithms,
none of which seems to be the {\it best} to be chosen.]
Anyway, if we want variations of entropy to be defined we also have to ensure some regularity condition
on the way the representative $\calU(L,\xi,\si)$ for the superpotential depends on the section $\si$.

\ss
Provided that all these {\it details} can be fixed,
now first evolution law of black holes has to be derived;
it establishes a relation between variations of the entropy, the mass
and angular momentum.
This is of course tricky because entropy is computed integrating on
$\Si$ while mass and angular momentum are computed on spatial
infinity.
Anyway, one can use the following property which holds
if $\Lie_\xi\si=0$, $\si$ is a solution of field equations and
$X$ is a [global] solution of the linearized field equations:
$$
\d[\de_X \calU(L,\xi,\si)-i_\xi\><\F(L,\ga)\circ j^{2k-1}\si\>\vert\>j^{k-1}X>]=0
\fl{\NMainTheorem}
$$\Compare{\NMainTheorem}{\MainTheorem}

Eq. $\MainTheorem$ integrated on $M$ gives
$$
\int_{\infty-\Si}[\de_X \calU(L,\xi,\si)-i_\xi\><\F(L,\ga)\circ
j^{2k-1}\si\>\vert\>j^{k-1}X>]=0
\fl{\NIntMT}
$$\Compare{\IntMT}{\NIntMT}                                                                

Now $\xi=t+\Om_{_{\rm H}}\la$
and $i_\xi\><\F(L,\ga)\>\vert\>j^{k-1}X>$ (which is linear in $\xi$) vanish on $\Si$;
thus it is easy to show that:
$$
\eqalign{
{\ka\over 2\pi}\de_X\calS=&\int_\Si \de_X\calU(L,\xi,\si)=\cr
=&\int_\Si [\de_X\calU(L,\xi,\si)-i_\xi\><\F(L,\ga)\circ j^{2k-1}\si\>\vert\>j^{k-1}X>]=\cr
=&\int_\infty [\de_X\calU(L,\xi,\si)-i_\xi\><\F(L,\ga)\circ j^{2k-1}\si\>\vert\>j^{k-1}X>]+\cr
-&\int_{\infty-\Si} [\de_X\calU(L,\xi,\si)-i_\xi\><\F(L,\ga)\circ
j^{2k-1}\si\>\vert\>j^{k-1}X>]=\cr =&
\int_\infty \de_X[\calU(L,\xi,\si)-\bfB(L,\xi,\si)]
= \de_X\calM^{_{\rm ADM}}-\Om\de_X\calJ^{_{\rm ADM}}\cr
\then& \de_X\calM^{_{\rm ADM}}={\ka\over 2\pi}\de_X\calS+\Om\de_X\calJ^{_{\rm ADM}}
\cr}
\fl{\FirstLaw}
$$
which is in fact the first law of thermodynamics.
Again $\calM^{_{\rm ADM}}$ and $\calJ^{_{\rm ADM}}$ are a generalization of ADM conserved
quantities $\ADMDefinition$ which reproduce the standard ones in standard General Relativity
when $\calU(L,\xi,\si)=\calU_{_{\rm Kom}}(L,\xi,\si)$ and
$\bfB(L,\xi,\si)=\bfB_{_{\rm ADM}}(L,\xi,\si)$. 
We stress that it is essential for this result that the term
$i_\xi\><\F(L,\ga)\>\vert\>j^{k-1}X>$ vanishes on $\Si$.
Under these hypotheses, eq.\ $\EntropyDef$ provides an expression for the variation of the entropy.
If one wants to have an expression for the entropy itself, one has to look for a new
quantity $\tilde \calU(L,\xi,\si)$ satisfing the variational equation
$$
\de_X \calU(L,\xi,\si)=\ka \de_X \tilde\calU(L,\xi,\si)
\fl{\NNewPot}$$\Compare{\NNewPot}{\NewPot}
so that one has
$$
\calS=2\pi \int_\Si \tilde\calU(L,\xi,\si)
\fn$$
We stress that the vanishing of $i_\xi\><\F(L,\ga)\>\vert\>j^{k-1}X>$
on $\Si$ is completely useless to derive $\FirstLaw$, if one defines the variation of the
entropy, instead of using $\EntropyDef$, by means of the following
$$
{\ka\over 2\pi}\de_X\calS=
\int_{\Si'} [\de_X\calU(L,\xi,\si)-i_\xi\><\F(L,\ga)\circ j^{2k-1}\si\>\vert\> j^{k-1}X>]
\fl{\NCorrectEntropy}$$\Compare{\CorrectEntropy}{\NCorrectEntropy}

Accordingly, the evaluation on $\Si$ is a further step needed to simplify expression
$\CorrectEntropy$. In fact, the integral
$$
\int_{\Si'} [\de_X\calU(L,\xi,\si)-i_\xi\><\F(L,\ga)\circ j^{2k-1}\si\>\vert\> j^{k-1}X>]
\fn$$
is completely unaffected by deformations of the region $\Si'$ because of $\MainTheorem$.

\NewSection{Acknoledgements}

One of us (L.F.) is grateful to R.\ Mann for having addressed our attention
to the example analysed in Section $8$ and
to R.\ McLenaghan and I.\ Volovich for fruitful discussions.

\NewSection{References}

\Biblio

\end